# All-fiber generation of 29-fs pulses at 1.3-μm via Cherenkov radiation

Luo Hao, Li Zhan*, Zhiqiang Wang, Liang Zhang and Cheng Feng

*State Key Laboratory of Advanced Optical Communication Systems and Networks, Key Laboratory for Laser Plasmas (Ministry of Education), Department of Physics and Astronomy, Shanghai Jiao Tong University, Shanghai 200240, China*
*Corresponding author: lizhan@sjtu.edu.cn*

**We have experimentally demonstrated the all-fiber generation of 1.3-μm femtosecond pulses via Cherenkov radiation (CR) from an ultrafast Er-doped fiber laser. The experiment shows that the pulses maintain below 40 fs in the range from 1270 to 1315 nm with multi-milliwatt average output power. The shortest generated pulses can be as short as 29 fs. This ultrashort pulse source in 1.3-μm window can bring the benefits to many fields such as the bio-imaging and the ultrafast spectroscopy, and so on.** © 2015 Optical Society of America

http:*********************************************

Ultrashort pulse sources at 1.3-μm band are very attractive for many applications such as optical communication and bio-imaging. Especially for bio-imaging, a lot of attention is focused on fluorescent imaging in the second near-infrared window (1–1.4 μm) due to minimal autofluorescence and tissue scattering [1-3]. The 1.3-μm excitation has superior penetration depth and much less phototoxic in nonlinear light microscopy (NLM) [4-7]. Traditionally, the main ultrashort pulse sources used for 1.3-μm NLM are the solid-state lasers including mode-locked Cr: forsterite laser and Ti: S-pumped optical parametric oscillator. Compared with traditional solid-state lasers, ultrafast fiber lasers offer more attractive features because of their cost efficiency, compactness, and ease of use.

Great efforts have been made to develop 1.3-μm ultrafast fiber lasers in the past decades. Sugawa *et al.* reported the 1.6-ps pulses generation from a 1.3-μm Pr-doped fluoride fiber laser[8]. Then, a 620-fs mode-locked Pr-doped fluoride fiber laser was demonstrated by M. J. Guy and co-workers [9]. In addition to Pr-doped fiber lasers, a 1.3-μm mode-locked bismuth fiber laser operating in both anomalous and normal dispersion regimes was presented [10]. The output pulses could be compressed down to 580 fs. However, shorter pulses are difficult to be got in 1.3-μm fiber lasers because of the relatively low gain of doped fibers, which demands long cavities and requires great efforts on dispersion management. Except for mode-locking technology, increasing efforts have also been carried on 1.3-μm pulses source through nonlinear frequency conversion technologies including soliton self-frequency shift (SSFS) and Cherenkov radiation (CR). Recently, Agrawal's group acquired a 227-fs pulse source via SSFS in a photonic crystal fiber [11]. Through Cherenkov radiation in highly nonlinear dispersion-shifted fiber, a 1.3-μm pulse source based on the system with free-space optical coupling was demonstrated [12]. By controlling the seeding pulse chirp with prism pairs, the output pulse wavelength could be tuned. The pulses as short as 24 fs were achieved with an optimized prism sequence. However, sub-100-fs 1.3-μm pulse sources based on an all-fiber configuration have seldom been reported up to date, and are still expected for biological and biomedical imaging as discussed in [13].

In this Letter we present an all-fiber, 1.3-μm femtosecond pulse source based on an Er-fiber ultrafast laser. The generated CR pulses can be as short as 29 fs. A desired output wavelength can be chosen in 1.3-μm window by adjusting the pump power. At the same time, the output pulse width is kept below 40 fs over the tuning range from 1270 to 1315 nm.

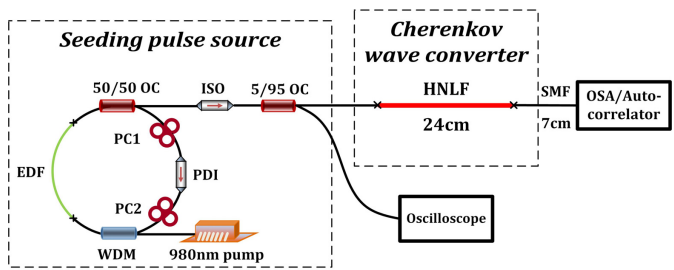

Fig.1. Experimental setup of 1.3-μm femtosecond pulse source.

As shown in Figure 1, the experimental setup consists of two parts: (i) an ultrafast Er-doped fiber laser used as a seeding pulse source (ii) a piece of highly nonlinear fiber (HNLF) used as a Cherenkov wave converter. The oscillator is a typical stretched-pulse fiber laser[14, 15].The gain fiber of the ring cavity is a 139 cm Er-doped fiber (OFS EDF80), forward pumped by a 980 nm laser diode through a 980/1550 wavelength division multiplexer (WDM). A polarization dependent isolator (PDI) sandwiched with two fiber polarization controllers (PC1 and PC2) is used as the mode-locking component in the cavity. A 50/50 optical coupler (OC) is located after the EDF to output the pulses. The total length of the cavity is 4.75 m, which corresponds to a repetition frequency rate of 42.4 MHz. The net dispersion of the cavity is +0.019 ps² at 1550nm. Positively chirped output pulses from the cavity can be compressed through optimizing the length of SMF outside the cavity. An optical isolator is used to remove backwards light. The output pulses are launched into a 95/5 coupler. The 5% output port is used to monitor the seeding pulses through an oscilloscope (RTO 1002, 2 GHz). The 95% OC port is linked to a 24cm long HNLF. The zero-dispersion wavelength of the used HNLF is 1522nm with a dispersion slope of 0.0165 ps/nm/km at 1550 nm. The mode field diameter (MFD) at 1550nm is 4.05 μm with a 10.1 $W^{-1}Km^{-1}$ nonlinear coefficient value. By splicing a 7cm long single-mode fiber jumper together with the HNLF end, we can measure the output pulses spectrum and pulse width with an optical spectrum analyzer (Yokogawa AQ6370C) and a second order autocorrelator (FR-103 XL).

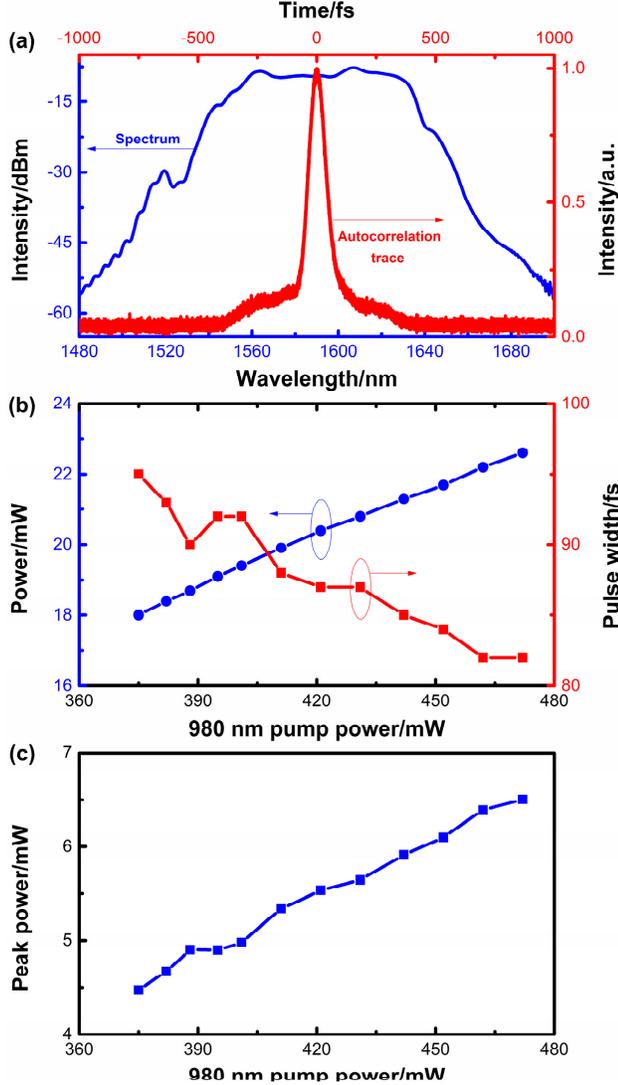

Fig.2. (a) Typical spectrum (blue lines) and autocorrelation trace (red lines) of the seeding pulses. (b)Seeding pulses average power (blue squares) and pulse width (red circles) vs the 980 nm pump power. (c) Seeding pulses peak power vs 980 nm pump power.

Before injecting the pulses into the HNLF, we firstly studied the features of the seeding pulses. By properly adjusting the polarization controllers, we achieved mode-locking by means of nonlinear polarization rotation [16, 17].The typical spectrum and autocorrelation trace of the compressed pulses are shown in Fig. 2(a). The central wavelength is 1592nm with a 3 dB bandwidth of 70nm and the pulse width is 92 fs. In Fig. 2(b), when the 980nm pump power increases from 375 to 472 mW, the output power linearly increases from 18 to 22.6mW, whereas the pulse width decreases from 95 to 82 fs. Through measuring the seeding pulse average power and pulse width, the peak power of the seeding pulses can be calculated. The peak power increases nearly linearly from 4.5 to 6.5 kW by adjusting the 980 nm pump power as shown in Fig. 2 (c). Thus, we can tune the seeding pulse peak power by adjusting the 980 nm pump power.

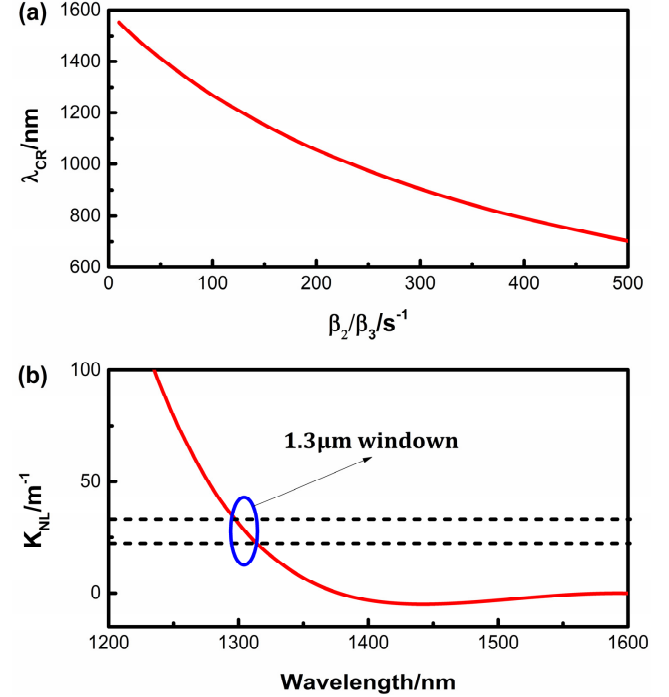

Fig.3. (a) Phase-matching wavelength of CR variance with the value of $\beta_2/\beta_3$ when the nonlinear term is neglected. (b) Phase-matching wavelength of CR variance with the nonlinear term.

Cherenkov radiation, also known as dispersive wave and nonsolitonic radiation, describes the energy transfer from a soliton to weak linear wave. It originates from a soliton perturbed by high order dispersion. The precise CR frequency is determined by phase-matching condition between the soliton and CR [18-21]. The seeding pulses propagating in the HNLF initially undergo higher-order soliton compression due to the combination of self-phase modulation (SPM) and anomalous dispersion. Around the maximal compression point, the CR pulses are generated. The CR frequency is determined by phase-matching condition and the formula is described below:

$$\frac{1}{2}\beta_2(\omega_{CR}-\omega_0)^2 + \frac{1}{6}\beta_3(\omega_{CR}-\omega_0)^3 = K_{NL} \quad (1)$$

where $\omega_{CR}$ and $\omega_0$ are the CR and input pulse central frequency, $\beta_2$ and $\beta_3$ are the second and the third dispersion coefficient at input central frequency, and $K_{NL}=1/2\gamma P_0$ is the nonlinear term ($\gamma$ is the nonlinear coefficient and $P_0$ is the peak power of input pulses)[21]. For simplicity, the higher-order dispersion term and Raman effect are not included.

Phase-matching curve of CR in HNLF is shown in Fig.3. The central wavelength of input pulses is 1592 nm. We report the phase-matched wavelength as a function of the value $\beta_2/\beta_3$ in Fig. 3(a).The nonlinear term is neglected here. It can be easily observed that by increasing the value $\beta_2/\beta_3$, the phase-matching condition is satisfied at shorter wavelength. With the aid of this phase-matching curve, we can choose a suitable HNLF to generate 1.3-μm CR. For the HNLF used in the experiment, $\beta_2$ is $2\times10^{-3}\ ps^2/m$ and $\beta_3$ is $3.25\times10^{-5}\ ps^3/m$. Thus the value $\beta_2/\beta_3$ is 61.5. The corresponding phase-matching wavelength is 1370 nm. Nevertheless, the nonlinear term has significant influence on CR generation as discussed below. In Fig. 3(b) we have studied the phase-matched wavelength of CR variance with the nonlinear term. It can be seen that when the value of $K_{NL}$ increases from 22.7 to 32.8 (namely, the pulse peak power increases from 4.5 to 6.5 kW), the CR wavelength blue shifts in 1.3-μm range. We can conclude that the phase-matched wavelength is a consequence of input pulses features (central wavelength and peak power) and dispersion parameter of HNLF (mainly the second and third dispersion).

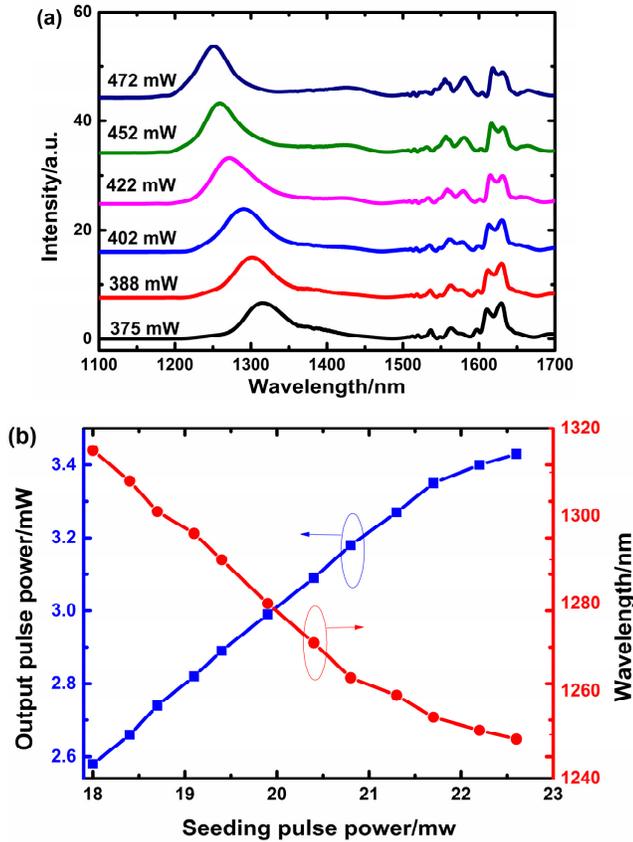

Fig.4. (a)Evolution of output spectrum in HNLF with increased 980 nm pump power (b)Output CR pulse power(blue squares) and CR central wavelength (red circles) vs the seeding pulse power.

The measured CR spectrum at different pump power is shown in Fig. 4(a). When the 980 nm pump power increases (namely, the increase of the seeding pulses peak power), the CR wavelength blue shifts in 1.3-μm window. This is in accord with theory description in Fig. 3(b). In the long-wave spectral component between 1.5 and 1.7 μm is residual seeding pulse spectrum. The emitted CR wavelength and power varying with increased seeding pulses power is shown as blue squares and red circles in Fig. 4(b), respectively. The central wavelength decrease nearly linearly from 1315 to 1249 nm. The output pulse power increases nearly linearly with a 15% conversion efficiency accounting for input seeding pulses power, and 1% conversion efficiency accounting for the 980 nm pump power. The output pulse power ranges from 2.6 to 3.4 mw corresponding to a single pulse energy of 58-76 pJ. This amounts to the single pulse energy of Pr-doped and Bi-doped mode-locked fiber laser [9, 10].

The autocorrelation trace of the generated CR pulses is measured at the end of 7 cm SMF. By carefully adjusting the phase matching angle of crystal in autocorrelator, we can get a sharp autocorrelation signal. The shortest AC trace is shown as black lines in Fig. 5(a).The 3dB width of the AC trace is 44 fs with no pedestal. Assuming that the CR pulses have a sech-form profile, the autocorrelation trace gives a pulse width of 29 fs. In Fig. 5(a), we use Gaussian and sech^2 function to fit the autocorrelation trace. It can be observed that the sech^2 function accords better with the experimental result. The central wavelength of the corresponding optical spectrum is 1296 nm with a 3 dB bandwidth of 65 nm shown in Fig. 5(b). The time–bandwidth product of the pulses is 0.32, suggesting that the pulses are almost transform limited.

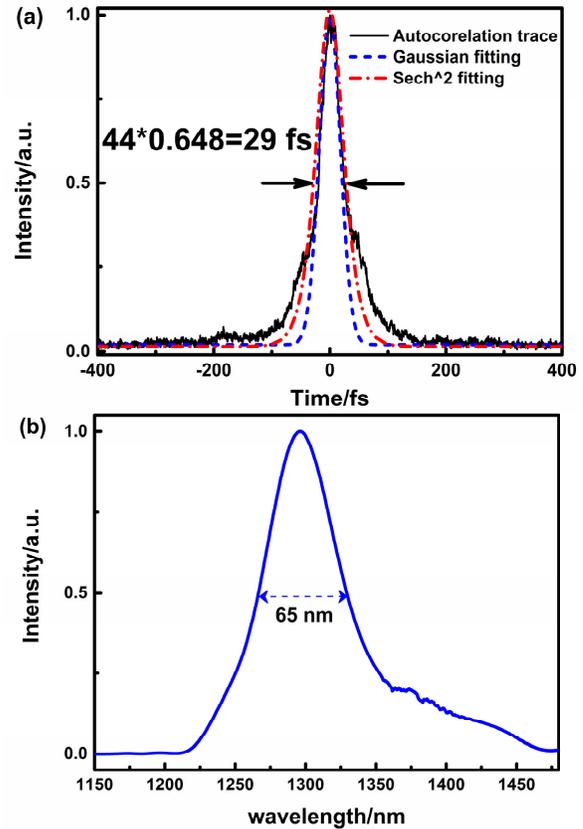

Fig. 5. (a) Autocorrelation trace and (b) optical spectrum of CR pulses.

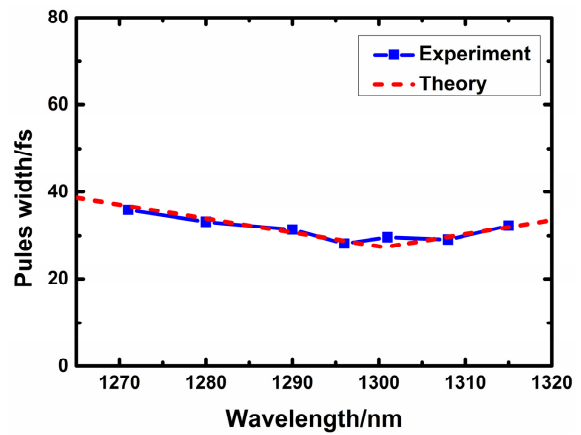

Fig.6. Experimental result (blue squares) and Theory result (red dash) of output CR pulse width vs central wavelength.

Also, we measured the output pulse width at different 980 nm pump power. As shown in Fig. 4, the generated CR wavelength can be tuned by adjusting 980 nm pump power. In Fig. 6, we studied the output pulse width varying with different CR central wavelength, which is controlled by adjusting 980 nm pump power. It can be observed that pulse duration below 40 fs can be maintained over a tuning range from 1270 to 1315 nm. The variation may be primarily owing to the dispersion of 7-cm SMF. Accounting for SMF dispersion value, a theory fitting curve is shown as red dash in Fig. 6. The theory results accord quite well with the experiment results. When the CR central wavelength is away from the SMF zero dispersion point, the output pulse width will be broadened.

In conclusion, we have demonstrated a 1.3-μm CR pulse source with an all-fiber configuration. The shortest pulse can be 29 fs. By simply adjusting 980 nm pump power, the emitted CR pulse wavelength can be tuned in the spectral range between 1270 to 1315 nm. At the same time, the pulse width can be maintained below 40 fs. We believe that our all-fiber, stable ultrashort pulse source in 1.3-μm window makes it promising for applications including ultrafast spectroscopy, optical coherence tomography and nonlinear light microscopy.

**Funding.** National Natural Science Foundation of China (NSFC) (61178014).